\documentclass[aps,pra,twocolumn]{revtex4-1}

\usepackage{graphicx}
\usepackage{amsmath}
\usepackage{amssymb}
\usepackage{dsfont}
\usepackage{url}
\usepackage[inline]{enumitem}
\usepackage{array}
\usepackage{multirow}
\usepackage{hyperref}

\newcommand{\ket}[1]{\left\vert #1\right\rangle }

\newcommand{\xpct}[1]{\left\langle #1\right\rangle }

\begin{document}

\title{Pairing Correlations and Separability Conditions in Identical Particle Systems}

\author{\c{C}. Aksak}
\email{e127474@metu.edu.tr}

\author{S. Turgut}
\affiliation{Department of Physics, Middle East Technical University, 06800, Ankara, Turkey}

\date{\today}
\begin{abstract}
	
Quantum correlations and entanglement in identical-particle systems have been a puzzling question which has attracted vast interest and widely different approaches. A novel approach is introduced by Kraus \emph{et al.}, [Phys. Rev. A \textbf{79}, 012306 (2009)] based on pairing correlations in fermionic systems and the use of witness formalism to detect pairing. In this contribution, this approach has been extended to bosonic systems and separability bounds based on pairing correlations for fermions and bosons have been obtained. A two-particle annihilation operator is used for constructing a two-particle observable as a candidate witness. Two different types of separability definition is introduced for bosonic systems and the separability bounds associated with each type are discussed. 
	
\end{abstract}
\maketitle

\section{Introduction}

Quantum theory allows correlations that cannot be described classically. Bell-type correlations were the first to be understood as unique quantum correlations that have no classical counterpart \cite{BrunnerBellnonlocality2014a}. By invalidating local hidden-variable theories, Bell correlations show that quantum theory is not a locally realistic theory, i.e., displays nonlocal correlations. A large class of entangled states exhibit nonlocal correlations, for instance; pure bipartite entangled states of any number of particles \cite{GisinMaximalviolationBell1992,PopescuGenericquantumnonlocality1992}. Despite this relation between entangled states and nonlocality, these two concepts are not synonymous: there are entangled states that do not have nonlocal correlations \cite{WernerQuantumstatesEinsteinPodolskyRosen1989a} and nonlocality can be observed without entanglement \cite{BennettQuantumnonlocalityentanglement1999}. From quantum information theoretical point of view both entanglement and nonlocality are viewed as resources for quantum information tasks, albeit of different kinds \cite{MethotAnomalyNonlocality2007}. In some quantum information processing scenarios such as teleportation, quantum key distribution protocols and quantum communication tasks, entanglement is interpreted as a resource shared between spatially separated systems. In these scenarios, there are two or more distinguishable particles (or systems) which are entangled or correlated through a quantum channel. In all these tasks distinguishability of the shared systems allows local agents to use local operations and classical communications (LOCC) to modify the entangled states for the information processing tasks they are trying to achieve. Since each system is locally accessible, the fermionic or bosonic nature of particles and issues associated with identical particles are irrelevant and effectively these particles are distinguishable \cite{SchliemannQuantumcorrelationstwofermion2001}. 

The advances in entanglement theory \cite{HorodeckiQuantumentanglement2009}, entanglement measures \cite{PlenioIntroductionEntanglementMeasures2007} and detecting entanglement \cite{GuhneEntanglementdetection2009} have found widespread applications in many-body and condensed-matter systems \cite{AmicoEntanglementmanybodysystems2008}. 
In many-body systems, particle-exchange symmetry necessitates the use of an antisymmetric or a symmetric wavefunction. Even for uncorrelated particles, use of Slater determinants for fermions and permanents for bosons create an impression of entanglement. However, when the same state is represented by occupation number formalism in the Fock space this impression is lost \cite{WisemanEntanglementIndistinguishableParticles2003}. According to one approach, this impression is a by-product of the use of quantum statistics and obscures the definition of entanglement and this can be avoided by making the observation that this appearance of entanglement is not real since indistinguishable particles are not addressable \cite{ZanardiQuantumentanglementfermionic2002,LiEntanglementtwoidenticalparticlesystem2001,ShiQuantumentanglementidentical2003,GhirardiGeneralcriterionentanglement2004,SchliemannQuantumcorrelationstwofermion2001,EckertQuantumCorrelationsSystems2002,ZanardiVirtualQuantumSubsystems2001,ZanardiQuantumTensorProduct2004,PaskauskasQuantumcorrelationstwoboson2001}. Following this conclusion, there are a number of proposals to resolve the question of how to define and quantify entanglement in identical-particle systems, which can be categorized into two groups:
\begin{enumerate*}[label=(\roman*)]
	\item Quantum Correlations \cite{EckertQuantumCorrelationsSystems2002,LiEntanglementtwoidenticalparticlesystem2001,PaskauskasQuantumcorrelationstwoboson2001,SchliemannQuantumcorrelationstwofermion2001,ShiQuantumentanglementidentical2003} and
	\item Entanglement of modes \cite{ZanardiQuantumTensorProduct2004,ZanardiQuantumentanglementfermionic2002,ZanardiVirtualQuantumSubsystems2001,GittingsDescribingmixedspinspace2002}
\end{enumerate*} 

\begin{enumerate*}[label=(\roman*)]
	\item  In Ref.~\onlinecite{SchliemannQuantumcorrelationstwofermion2001} Schliemann \emph{et al.} propose a Slater-rank criterion for quantum correlations of fermions. An extension of this work to bosons can be found in \cite{PaskauskasQuantumcorrelationstwoboson2001}. \item In \cite{ZanardiVirtualQuantumSubsystems2001,ZanardiQuantumTensorProduct2004}, Zanardi \emph{et al.} claim that entanglement is relative to a tensor product structure based on partitioning (modes) of the system and the partitioning, naturally, relies on the locally accessible observables. Therefore, entanglement is observable induced.
\end{enumerate*}  

In \cite{WisemanEntanglementIndistinguishableParticles2003} Wiseman and Vacarro compare and review the two different approaches. They argue that, the quantum correlations of \cite{PaskauskasQuantumcorrelationstwoboson2001} give different results for fermions and bosons which are in the same bipartite state in mode-occupation representation. Also, note that quantum correlations do not comply with the LOCC paradigm. Wiseman and Vacarro continue that entanglement of modes are only meaningful if particle number conservation is taken into account when partitioning identical particles into spatially distinguishable parties. 

A completely different approach is introduced in \cite{IchikawaExchangesymmetrymultipartite2008}, where authors argue that exchange symmetry causes indistinguishable particles to be strongly entangled. There are proposals of extracting this kind of entanglement and registering onto spatially distinguishable modes, aiming to access entanglement coming from (anti)symmetrization  \cite{KilloranExtractingEntanglementIdentical2014, CavalcantiUsefulentanglementPauli2007, OmarSpinspaceentanglementtransfer2002}.   

Whichever definition of entanglement is employed, the associated separable states, which are defined as mixtures of unentangled states, form a convex subset $\mathcal{S}$ of all possible states. This convex set can be equivalently described by the set of its tangent hyperplanes. Such a description is usually employed with the witness formalism\cite{HorodeckiSeparabilitymixedstates1996,*Terhalfamilyindecomposablepositive2001}. An entanglement witness is a Hermitian operator $W$ that has non-negative expectation values in all separable states and a negative expectation value in some state, i.e.,
\begin{align*}
 &(i)  & \xpct{W}=\mathrm{tr}\left(\rho_{sep}W\right)\geq0 \qquad \forall\,\rho_{sep}\in\mathcal{S}\\
 &(ii) & \xpct{W}=\mathrm{tr}\left(\rho W\right)<0 \qquad \exists\,\rho\notin\mathcal{S} ~~~~
\end{align*}
Expressed differently, this formalism describes entanglement and separability by using the expectation values of observables. 

Kraus \emph{et al.} \cite{KrausPairingfermionicsystems2009} proposed using the witness formalism for the description of particle correlations in many-fermion systems. They define the product states as states with a definite number $N$ of fermions and having the form
\begin{equation}
  \ket{\Psi}= c^\dagger_1 c^\dagger_2\cdots c^\dagger_N\ket{0}~,
  \label{eq:productState}
\end{equation}
where $c_i^\dagger$ are fermion creation operators and $\ket{0}$ is the vacuum state. In other words, product states are states with Slater rank 1, which corresponds to unentangled states of Schliemann \emph{et. al} \cite{SchliemannQuantumcorrelationstwofermion2001} . A mixed state is called separable if it can be expressed as a mixture of product states. Finally, a correlated many-fermion state is any state which is not separable. The characteristic property of the correlated states is of course the expectation values of some observables that somehow incorporate the correlations between the particles, which are distinct from the corresponding expectation values in separable states in some way. The witness formalism enables us to qualify and quantify that distinction and therefore can be used for detecting particle correlations.

It appears that, expectation values of single-particle operators, i.e., operators of the form $\sum B_{i,j}c_i^\dagger c_j$, cannot be used as a witness. At least two-particle operators of the form $\sum B'_{ij,k\ell}c_i^\dagger c_j^\dagger c_\ell c_k$, perhaps in conjunction with single-particle operators, are needed to build a witness. Kraus \emph{et al.} then define a particular kind of correlation, the pairing correlation, as the one that can be detected by a two-particle observable. Pairing does not capture all possible many-body correlations among fermions because there are correlated states which do not display pairing correlations. Kraus \emph{et al.} then propose and discuss several two-particle witnesses and measures of pairing correlations in fermionic systems.
 
The purpose of this contribution is to use the same approach in quantifying pairing correlations in bosonic systems as well. In Section II, the basic witness operator that will be used is introduced. After a brief review of fermionic case, bosonic product states are defined and the formalism is applied to bosons. Finally, a brief conclusion is given.

\section{Pairing Witness}

Out of infinitely many possible pairing witnesses, some choice has to be made based on its amenability to analytic treatment. For this purpose, this article concentrates on observables of the form $Q^\dagger Q$, where $Q$ is a 2-particle annihilation operator, which can in general be written as
\[
  Q=\frac{1}{2} \sum_{ij} A_{ij} c_i c_j~,
\]
where $A$ is some complex square matrix and $c_i$ are annihilation operators corresponding to orthonormal single-particle states. The matrix $A$ should be anti-symmetric for fermions and symmetric for bosons. It is possible to simplify the form of $Q$ by using passive transformations of the form $c_i \longrightarrow c'_i=\sum_j (U^{-1})_{ij}c_j$ where $U$ is unitary. In such a case, the matrix $A$ transforms as $A\longrightarrow A'=U^TAU$, where superscript $T$ represents matrix transposition, i.e., $A$ changes by unitary congruence. By a suitable choice of $U$, it is possible to choose single-particle states such that (i) $A$ is block diagonal with $1\times1$ zero or $2\times2$ anti-symmetric blocks for the fermionic case and (ii) $A$ is diagonal for the bosonic case\cite{HornMatrixAnalysis2012a}. In addition to this, the matrix elements of $A$ can be made real. It will be assumed that such a redefinition has been made. In other words, in the most general case, the $Q$ operator can be expressed as
\begin{align}
 Q &= A_1 c_1 c_2 + A_2 c_3 c_4 +\cdots + A_rc_{2r-1}c_{2r} ~~\textrm{(fermions)}~,
 \label{generalQforFermions}\\
 Q &= \frac{1}{2}(A_1 c^2_1  + A_2 c^2_2 +\cdots + A_rc^2_{r}) ~~\textrm{(bosons)}~,
	\label{generalQforBosons}
\end{align}
where $A_1,\ldots,A_r$ are positive real numbers. The number of terms $r$ will be called as the \emph{rank} of $Q$. It is necessary for detecting pairing correlations that $r\geq2$. There is an arbitrariness in the choice of the single-particle states that enter into the definition of $Q$. But, once this choice is made, the \emph{standard basis} of single-particle states will be redefined such that $Q$ depends on the the first $2r~/~r$ states of this basis for fermions/bosons. The coefficients $A_i$ can in principle be different from each other. However, to be able to obtain closed-form analytical expressions, only the choice $A_1=A_2=\cdots=A_r=1$ is considered in this article.

Let $\Lambda^\mathrm{sep}_{r,N}$ be the largest expectation value of $Q^\dagger Q$ for separable states and for rank $r$ operator $Q$, i.e.,
\begin{equation}
  \Lambda^\mathrm{sep}_{r,N} =\sup_{\rho_\mathrm{sep}} \mathrm{tr}\, \rho_\mathrm{sep} Q^\dagger Q = \sup_{\ket{\Psi_\textrm{prod}}} \langle\Psi_\textrm{prod}\vert Q^\dagger Q \vert\Psi_\textrm{prod}\rangle ~,
\end{equation}
where the supremum is taken over the separable states $\rho_\mathrm{sep}$, or equivalently it is taken over the pure product states $\ket{\Psi_\textrm{prod}}$. After that, the observable 
\begin{equation}
  W=  \Lambda^\mathrm{sep}_{r,N}\mathds{1} - Q^\dagger Q
  \label{eq:general_witness}
\end{equation}
can be a possible candidate for a witness. A negative expectation value of $W$, in other words, satisfaction of the $\langle Q^\dagger Q\rangle> \Lambda^\mathrm{sep}_{r,N}$ inequality, indicates a correlated state. Let $\lambda_{r,N}$ denote the maximum eigenvalue of $Q^\dagger Q$. If $\lambda_{r,N}> \Lambda^\mathrm{sep}_{r,N}$ holds, then $W$ can be used as a witness for some correlated states.

\subsection{Fermionic Case}

One of the possible pairing tests discussed by Kraus \emph{et al.} is quite similar to the witness in Eq.~\eqref{eq:general_witness}. For this reason, only a brief summary and a few extra details will be given in here. Let $c_i$ represent the annihilation operators for normalized, mutually orthogonal single-particle states (the standard basis states). Among these states, the first $2r$ orbitals are chosen to define the $Q$ operator by
\begin{equation}
  Q=\sum_{i=1}^r c_{2i-1} c_{2i} =c_1c_2+c_3c_4+\cdots+c_{2r-1}c_{2r}~.
\end{equation}
For the product state formed by a subset of standard basis states
\begin{equation}
  \ket{\Psi}=c^\dagger_{k_1}c^\dagger_{k_2}\cdots c^\dagger_{k_N}\ket{0} ~,
\end{equation}
the expectation value $\langle Q^\dagger Q\rangle$ is equal to the number of $i$ ($1\leq i\leq r$) such that both $2i-1$ and $2i$ are occupied (i.e., both labels are in the list of occupied states $k_1,\ldots,k_N$.) For $N\geq2r$, the expectation value can be maximized by occupying all of the first $2r$ single-particle states, in which case $\langle Q^\dagger Q\rangle=r$. For the remaining cases of $N<2r$, it can easily be seen that the maximum is obtained by occupying the first $N$ orbitals consecutively. By using the technique used by Kraus \emph{et al.}, it can be easily shown that these are also the maxima over all separable states. Consequently, the separable-state maximum of $\langle Q^\dagger Q\rangle$ for $N$-fermion states is given by
\begin{equation}
 \Lambda^\mathrm{sep}_{r,N} = \left\{
    \begin{array}{ll}
      \lfloor\frac{N}{2}\rfloor & \textrm{for } N<2r~,\\
      r & \textrm{for } N\geq2r~,
    \end{array}
	\right.
\end{equation} 
where $\lfloor x\rfloor$ denotes the integer-part (floor) function.

For computing the maximum eigenvalues $\lambda_{r,N}$, it is useful to notice that SU(2) algebra relations, $[J_i,J_j]=i\sum_k\epsilon_{ijk}J_k$, are obeyed by
\begin{align}
J_x &=\frac{Q+Q^\dagger}{2}\\
J_y &=i\frac{Q-Q^\dagger}{2}\\
J_z &=\frac{N_Q-r}{2}
\end{align}
where
\begin{equation}
 N_Q=\sum_{i=1}^{2r} c_i^\dagger c_i
\end{equation}
is the number of particles in the first $2r$ orbitals. The corresponding ladder operators are $Q=J_x-iJ_y$ and $Q^\dagger=J_x+iJ_y$. By using the well-known angular momentum algebra, it is straightforward to find all eigenstates and eigenvalues of $Q^\dagger Q=J^2-J_z^2+J_z$. 
Therefore, if (i) $\ket{\psi_0}$ is a state annihilated by $Q$ and (ii) it is also an eigenstate of $N_Q$ with eigenvalue $\nu$, then $\nu\leq r$ should be satisfied and 
\begin{equation}
 \ket{\psi} = (Q^\dagger)^\mu \ket{\psi_0} \qquad (0\leq\mu\leq r-\nu)
\end{equation}
is an eigenstate of $Q^\dagger Q$ with eigenvalue $\mu(r-\nu-\mu+1)$. All eigenstates of $Q^\dagger Q$ can be obtained by this procedure. Finding the maximum eigenvalue of $Q^\dagger Q$ among $N$-fermion states is straightforward but tedious. The final result is
\begin{align}
  \lambda_{r,N} &=\max_{0\leq \mu  \leq r,N/2} \mu(r+1-\mu) \\
    &=\left\{
    \begin{array}{ll}
    	\lfloor\frac{N}{2}\rfloor (r+1-\lfloor\frac{N}{2}\rfloor) & \textrm{for } N< 2\lfloor \frac{r+1}{2}\rfloor~,\\
    	\lfloor \frac{r+1}{2}\rfloor\lceil\frac{r+1}{2}\rceil  & \textrm{for } N\geq 2\lfloor \frac{r+1}{2}\rfloor~,
    \end{array}
    \right.
\end{align}
where $\lceil x\rceil$ represents the ``ceiling'' function, the smallest integer greater than or equal to $x$.

For the two trivial cases of $N=1$, for which there can be no correlations, and $r=1$, for which $\langle Q^\dagger Q\rangle$ does not contain pairing information, $\lambda_{r,N}=\Lambda^\mathrm{sep}_{r,N}$ holds and therefore $W$  is not a witness. Apart from these, there is one non-trivial situation where the $Q$ operator has rank $r=2$ and the number of particles is $N\geq4$, for which case $\lambda_{2,N}=\Lambda^\mathrm{sep}_{2,N}=2$ holds. Consequently, the current approach cannot detect any pairing correlations for such a situation. It is interesting to see if a rank $r=2$ operator in the general form of Eq.~\eqref{generalQforFermions} with $A_1\neq A_2$ can function as a witness for $N\geq4$. It can be shown that, for $Q=A_1c_1c_2+A_2c_3c_4$, the maximum eigenvalue of $Q^\dagger Q$ is $A_1^2+A_2^2$ with the corresponding eigenstates being
\begin{align}
 \vert \Psi_1 \rangle & = c_4^\dagger c_3^\dagger c_2^\dagger c_1^\dagger 
  ~ 
  c^\dagger_{k_1} \cdots c^\dagger_{k_{N-4}}\vert0\rangle ~,
	\label{ProductForm}\\
 \vert \Psi_2 \rangle & = (A_1c_2^\dagger c_1^\dagger +A_2c_4^\dagger c_3^\dagger) c^\dagger_{k_1} \cdots c^\dagger_{k_{N-2}}\vert0\rangle ~,
\end{align}
where $k_\sigma\geq5$. Since the topmost form of the states in the above list are already product states, $\lambda_{2,N}=\Lambda_{2,N}^\mathrm{sep}$ equality keeps holding. The rank 2 case appears to be an important deficiency of the current approach.

For the cases $r>2$ and $N\geq2$, the strict inequality  $\lambda_{r,N}>\Lambda^\mathrm{sep}_{r,N}$ holds, and therefore the associated witness can be used to detect correlations. It can also be observed that, as the rank $r$ increases, the ratio $\lambda_{r,N}/\Lambda^\mathrm{sep}_{r,N}$ becomes larger without a bound. This means that large $r$ values should be preferred for overcoming the effects of measurement uncertainties.

\subsection{Bosonic Case}

It appears that there are two possible reasonable ways to implement the product-state definition in Eq.~\eqref{eq:productState} to bosons. Let $c_i$ ($i=1,2,\ldots, M$) be boson annihilation operators corresponding to an orthonormal set of $M$ single-particle states. For any arbitrary single-particle state $\phi=(\phi_1,\ldots,\phi_M)$, here considered as an $M$-dimensional (matrix) vector, the corresponding annihilation operator is defined as
\begin{equation}
 c(\phi)=\sum_{i=1}^M \phi_i^*c_i~. 
\end{equation}
The canonical commutation relations can then be expressed as
\begin{equation}
 [c(\phi),c^\dagger(\chi)] = \langle\phi\vert\chi\rangle=\sum_{i=1}^M \phi_i^*\chi_i
 \quad\textrm{ and }\quad
 [c(\phi),c(\chi)] =0~.
\end{equation}
An $N$-particle bosonic product state can be defined as a state of the form 
\begin{equation}
 \vert\Psi\rangle = (const) 
  c^\dagger(\phi_1)  c^\dagger(\phi_2) \cdots c^\dagger(\phi_N)\vert0\rangle 
\end{equation}
where $\phi_1, \phi_2,\ldots, \phi_N$ are some single-particle states and $(const)$ refers to a normalization constant. The following alternative definitions of product states impose different additional conditions on the single-particle states.
\begin{enumerate}
\item Type 1 product states are those $\vert\Psi\rangle$ where the single-particle states $\phi_1, \phi_2,\ldots, \phi_N$  are either mutually orthogonal or mutually identical. In this case, there is an orthonormal set of single particle orbitals $\alpha_1,\ldots,\alpha_p$ and the state is 
\begin{equation}
 \vert\Psi\rangle= \frac{(c^\dagger(\alpha_1))^{m_1}}{\sqrt{m_1!}} \cdots \frac{(c^\dagger(\alpha_p))^{m_p}}{\sqrt{m_p!}}\vert0\rangle
\label{eq:type1State}
\end{equation}
i.e., state $\alpha_b$ is occupied by $m_b$ bosons and there are $N=m_1+\cdots+m_p$ bosons.

\item Type 2 product states are those $\vert\Psi\rangle$ where the single-particle states $\phi_1, \phi_2,\ldots, \phi_N$ are not required to satisfy additional conditions, i.e., $\phi_i$ and $\phi_j$ are not required to be either identical or orthogonal and the overlap $\langle\phi_i\vert\phi_j\rangle$ can have any value. 
\end{enumerate}

Obviously, type 1 product states form a subset of type 2 states, and the same subset relation holds for the corresponding separable states. Unfortunately, the separable-state limit $\Lambda^\mathrm{sep}_{r,N}$ cannot be computed in a closed-form analytical expression for type 2 states. This is the principal reason for defining type 1 product states as an alternative definition. Below, analytical expressions for $\Lambda^\mathrm{sep}_{r,N}$ will be given for type 1 separable states and only a few numerical computations will be provided for type 2 states.

For bosons, the standard basis for single-particle states is assumed to be chosen such that the $Q$ operator appears as 
\begin{equation}
  Q=\frac{1}{2}\sum_{i=1}^r c_i^2~,
\end{equation}
In this case, the associated commutation relations are as follows,
\begin{align}
  [Q,Q^\dagger] &= N_Q+\frac{r}{2}~,\\
  [N_Q,Q^\dagger] &= 2Q^\dagger ~,
\end{align}
where 
\begin{equation}
  N_Q=\sum_{i=1}^r c_i^\dagger c_i
\end{equation}
is the number of bosons in the first $r$-states. Consequently, the three operators, $Q$, $Q^\dagger$ and $N_Q+r/2$ generate the Lie algebra SL(2,$\mathbb{R}$) associated with the special linear group of $2\times2$ real matrices. In parallel to the fermionic case, the operators $Q$ and $Q^\dagger$ act as ladder operators for the common eigenstates of $Q^\dagger Q$ and $N_Q$. 

All common eigenstates and the corresponding eigenvalues of $Q^\dagger Q$ and $N_Q$ can be found as follows. Let $\nu$ be any non-negative integer ($\nu=0,1,2,\ldots$). Let $f(x_1,x_2,\ldots,x_r)$ be $\nu$th degree homogeneous polynomial solution of $r$-dimensional Laplace's equation in $r$-dimensions,
\begin{equation}
\left(\frac{\partial^2}{\partial x_1^2}+\cdots+\frac{\partial^2}{\partial x_r^2}\right)f=0 ~.
\end{equation}
and let $g(x_{r+1},\cdots,x_M)$ be any homogeneous polynomial of degree $\nu'$. Then the state
\begin{equation}
 \vert\psi_0\rangle=f(c_1^\dagger,\ldots,c_r^\dagger)g(c_{r+1}^\dagger,\ldots)\vert0\rangle
\end{equation}
is a state which is annihilated by $Q$ (i.e., $Q\vert\psi_0\rangle=0$). This state is also an eigenstate of $N_Q$ with eigenvalue $\nu$ and the total number of bosons is $N=\nu+\nu'$. Let $\mu$ be another non-negative integer. Then
\begin{equation}
 \vert\psi\rangle=(Q^\dagger)^\mu\vert\psi_0\rangle
\end{equation}
is a state with $N=\nu+\nu'+2\mu$ bosons with $N_Q$ value equal to $\nu+2\mu$. It is straightforward to establish that the corresponding eigenvalue of $Q^\dagger Q$ is given by
\[
\mu(\nu+\frac{r}{2}+\mu-1)~.
\]
Using this, the maximum eigenvalue of $Q^\dagger Q$ among all $N$-particle states can be computed as
\begin{equation}
  \lambda_{r,N} =\left\{
 \begin{array}{ll}
 	\frac{N(N+r-2)}{4}& \textrm{if $N$ is even,} \\
 	\frac{(N-1)(N+r-1)}{4}& \textrm{if $N$ is odd.} \\
 \end{array}
 \right.
\end{equation}

\begin{figure}
\includegraphics[scale=0.6]{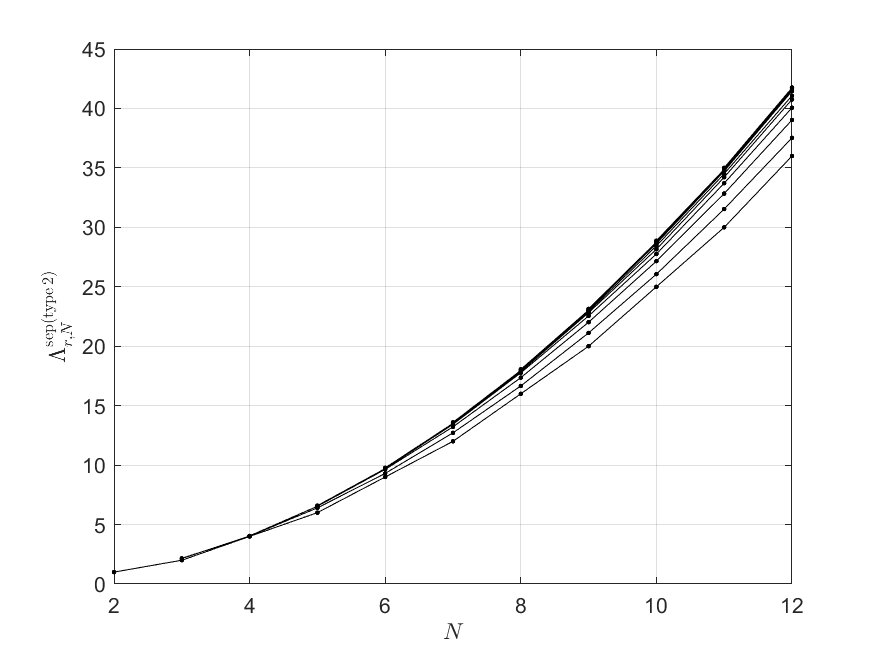}
\caption{Type 2 separable-state bound, i.e., maximum expectation value of $Q^\dagger Q$ in type 2 separable states, is plotted as a function of the total number of particles. The bound also depends on the rank $r$. The lowest curve is for $r=2$ and the value of $r$ changes by 1 between two successive curves. Note that the value of the bound for $r>N$ is identical with the bound for $r=N$.}
\label{fig1}
\end{figure}

In the Appendix, it is shown that the separable-state bound for type 1 separable states is given as
\begin{equation}
 \Lambda^\mathrm{sep (type\,1)}_{r,N} =\left\{\begin{array}{ll}
   \frac{N^2}{4} & \textrm{if } N  \textrm{ is even,} \\
   \frac{N^2-1}{4} & \textrm{if } N \textrm{ is odd.} 
 \end{array}\right.
 \label{eq:type1sep}
\end{equation}
Unfortunately there is no closed-form expression for the bound $\Lambda^\mathrm{sep (type\,2)}_{r,N}$ for type 2 separable states. Fig.~\ref{fig1} shows the numerically computed values of the type 2 bound.

\begin{figure}
\includegraphics[scale=0.6]{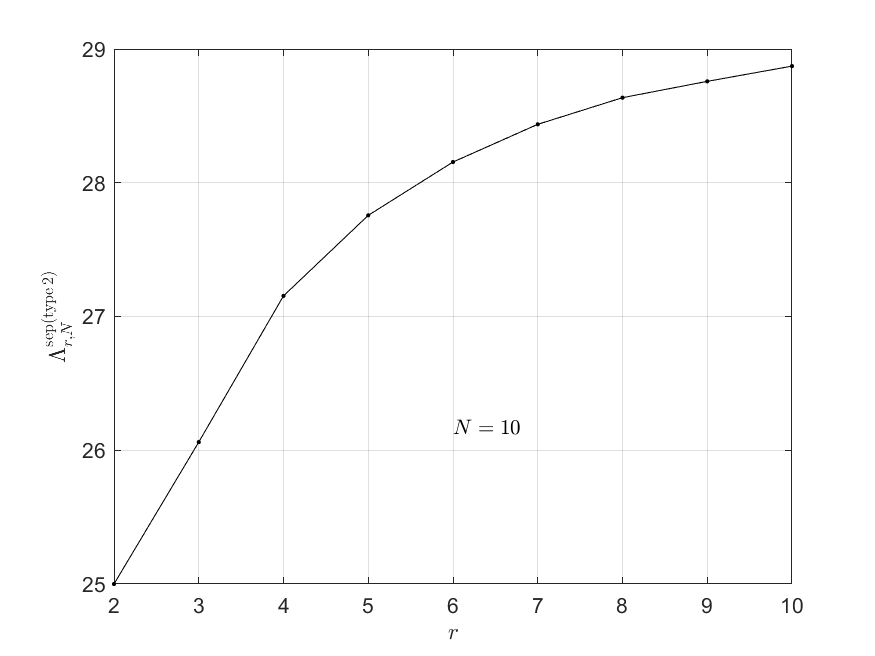}
\caption{Type 2 separable-state bound for $N=10$ particles for different values of the rank $r$. The value $r=2$ is equal to the type 1 bound. Also, note that the bound for $r>N$ is equal to the bound for $r=N$.}
\label{fig2}
\end{figure}

An interesting feature of the type 1 bound $\Lambda_{N,N}^\mathrm{sep(type\,1)}$ is its independence of the rank $r$. This is related to the fact that the product state of the form in Eq.~\eqref{eq:type1State} that maximizes the expectation value $\langle Q^\dagger Q\rangle$ has only $p=2$ single-particle states, both of which are inside the linear span of the first $r$ standard basis states. The value of $\langle Q^\dagger Q\rangle$ depends only on the overlap between states $\alpha_k$ and their complex conjugates. For this reason, the size of $r$ never enters into expectation values.

For type 2 states, the numerical results indicate a dependence on $r$. In fact, the set of single-particle states $\phi_1,\ldots,\phi_N$ that maximize $\langle Q^\dagger Q\rangle$ have different geometrical relations to each other which strongly depends on the value of $r$. In all cases, these $N$ single-particle states appeared within the subspace spanned by the first $r$ standard-basis states. This implies that, if the rank $r$ is larger than the number of particles, then the separable-state bound does not depend on $r$, i.e.,
\begin{equation}
  \Lambda_{r,N}^\mathrm{sep(type\,2)} = \Lambda_{N,N}^\mathrm{sep(type\,2)} \quad \textrm{for } r>N~.
\end{equation}
In the non-trivial range $2\leq r\leq N$, the bound $\Lambda_{r,N}^\mathrm{sep(type\,2)}$ appears to be monotonically increasing with $r$. The dependence of the bound on the rank for the special case of $N=10$ is shown on Fig.~\ref{fig2}. In all cases computed, $\Lambda_{r,N}^\mathrm{sep(type\,2)}$ appeared to be of the same order of magnitude as $\Lambda_{r,N}^\mathrm{sep(type\,1)}$.

It can be seen that, experimental detection of pairing correlations between identical bosons can be challenging because of the smallness of the size of the violation. The following can be checked for type 1 states,
\begin{equation}
 \frac{\lambda_{r,N}}{\Lambda_{r,N}^\mathrm{sep(type\,1)}} \sim \frac{N+r}{N}~,
\end{equation}
which continues to hold within the same order of magnitude for type 2 states. Consequently, for systems with a large number of particles, the margin of detection of pairing correlations is rather tight. One must either resort to high rank schemes $r\sim N$, or reduce measurement uncertainty.

\subsubsection{Rank 2 case}

Similar to the fermionic case, the special case of rank $r=2$ does not yield a valid witness as $\lambda_{2,N}=\Lambda_{2,N}^\mathrm{sep(type\,1)}=\Lambda_{2,N}^\mathrm{sep(type\,2)}$ holds. The equality of $\Lambda_{2,N}^\mathrm{sep(type\,2)}$ has only been verified numerically: It appears that the optimum single-particle states $\phi_1,\ldots,\phi_2$ that maximize $\langle Q^\dagger Q\rangle$ for type 2 states are dispersed into two orthogonal states. In other words, the optimum type 2 separable state is also a type 1 state. In addition to this, the maximum eigenvalue of $Q^\dagger Q$ also coincides with these bounds and hence there is no witness.

It is possible to do a similar calculation and see if this is an artifact of the assumption $A_1=A_2$ in the general form in Eq.~\eqref{generalQforBosons}. For this purpose consider
\begin{equation}
 Q=\frac{1}{2}\left(A_1c_1^2+A_2c_2^2\right)
\end{equation}
where $A_1\neq A_2$ and see if it is possible to get a non-trivial witness. Note that, eigenstates of $Q^\dagger Q$ can be expressed in the form,
\begin{equation}
 \vert\Psi\rangle=f(c_1^\dagger,c_2^\dagger) \, c_{k_1}^\dagger \cdots c_{k_m}^\dagger\vert0\rangle
\end{equation}
where $k_\sigma\geq3$ and $f(x_1,x_2)$ is a homogeneous polynomial of two variables. By the fundamental theorem of algebra, this polynomial can be written as a product of homogeneous polynomials of degree 1. Consequently, all eigenstates of $Q^\dagger Q$ can be described as type 2 product states and therefore
\begin{equation}
 \lambda_{2,N} = \Lambda_{2,N}^\mathrm{sep(type\,2)}~.
\end{equation}
This shows that, as far as type 2 separability is concerned, it is not possible to build a witness from $Q$ operators of rank 2. As type 1 separable states form a proper subset of type 2 states, it appears that type 1 witnesses are still possible with rank-2 $Q$ operators.

\section{Conclusions}

In conclusion, it is shown that by borrowing entanglement witness formalism, pair correlation measurements for bosonic many-particle systems can be performed, similar to \cite{KrausPairingfermionicsystems2009}. Separability bounds have been calculated for the observable \eqref{eq:general_witness} for two different kinds of separability; either the single-particle orbitals are from an orthonormal set, for type 1 separability, or they can have a non-zero overlap for type 2 separability. Type 2 separability reflects the fact that mixing two unpaired bosonic systems should not create pairing owing to the only fact that single particle orbitals have non-zero overlap. Type 1 separability is a special case of type 2 separability. A closed form analytical expression has been provided for separability bound for type 1 states in Eq.~\eqref{eq:type1sep}, however only numerical results are provided for type 2 states. A crucial observation on the difference between two types of separability is that there is no rank dependency for type 1. The differences between the two types of separability are summarized in the table below. 

\begin{center}
	\begin{tabular}{|m{4.5cm}|c|c|}	
		
		\hline
		 & \textbf{Type 1} & \textbf{Type 2} \\
		 \hline
		 Single particle orbitals are from an orthonormal set & Yes & No \\
		 \hline
		 Separability bound depends on the rank of Q & No & \begin{tabular}{c}
		 	for  $r<N$ Yes \\ \hline for $r\geq N$ No \end{tabular} \\
		 \hline
		 Witness at rank $r=2$ & No & No \\
		 \hline
	\end{tabular}
\end{center}

Pairing is a nonclassical effect which arises due to particle interactions and influenced by indistinguishability. Extracting entangled electrons from paired ones in superconductors has been demonstrated \cite{RecherAndreevtunnelingCoulomb2001,*SchindeleNearUnityCooperPair2012,*LesovikElectronicentanglementvicinity2001}. Recently, it is shown that spatially entangled bosons can be extracted out of Bose-Einstein condensates \cite{FadelSpatialentanglementpatterns2018,*KunkelSpatiallydistributedmultipartite2018,*LangeEntanglementtwospatially2018}.
Although, pairing and entanglement are not synonymous, they are interrelated and with a better understanding of the relationship between them, good entanglement sources might be better identified. An open question of interest is then the following: how are interaction sensitive pairing correlations related to the accessible entanglement in different interaction regimes in many-body systems, such as ultracold gases \cite{BlochManybodyphysicsultracold2008}?



\appendix*
\section{Separability bound for Type 1 states}
Let $\vert\Psi\rangle$ be the type 1 product state given in Eq.~\eqref{eq:type1State}. The expectation value of $Q^\dagger Q$ in this state is given by 
\begin{equation}
 \langle Q^\dagger Q\rangle_\Psi = \sum_{a<b} \vert R_{ab} \vert^2 m_a m_b + \sum_a \frac{1}{4} \vert R_{aa} \vert^2 m_a (m_a-1)~,
\end{equation}
where 
\begin{equation}
R_{a,b} =\sum_{i=1}^r \alpha_{a,i}\alpha_{b,i}~,
\label{eq:RDefined}
\end{equation}
and $m_a$ are positive integers that satisfy $\sum_am_a=N$. It is useful to define the following anti-linear operation for single-particle states:  $\phi=(\phi_1,\phi_2,\cdots)\longrightarrow\phi^*=(\phi_1^*,\phi_2^*,\ldots)$, which is simply complex conjugation in the standard basis. The state $\phi^*$ will be called as the conjugate of $\phi$. Let $P$ denote the projection to the subspace spanned by the first $r$ single-particle states. Then, $R_{ab}$ can be expressed as
$R_{ab}=R_{ba}=\langle \alpha_a^*\vert P\vert \alpha_b\rangle$, i.e., it is the overlap between $\alpha_a^*$ and $\alpha_b$ on the aforementioned subspace. It can be seen that $D_{ab}=\vert R_{ab}\vert^2$ is a $p\times p$, symmetric, doubly substochastic matrix, i.e.,
\[
 \sum_{b=1}^p D_{ab}\leq1~.
\]
At this point, it is useful to rewrite the expectation value in terms of $D$, $ \langle Q^\dagger Q\rangle_\Psi = f(D)$ where
\begin{equation}
 f(D) \equiv \frac{1}{2}\sum_{a\neq b}  D_{ab} m_a m_b + \frac{1}{4} \sum_a D_{aa} m_a (m_a-1)~,
 \label{eq:expValueInType1}
\end{equation}
and notice that the right-hand side is a linear function of the matrix elements of $D$ with non-negative coefficients. For this reason, it is easier to first maximize the above expression with respect to $D$ and then do the maximization with respect to the number of particles $m_a$.

However, the restriction $D_{ab}=\vert R_{ab}\vert^2$ together with Eq.~\eqref{eq:RDefined} makes the maximization problem complicated. To simplify this problem, the following approach is taken: First, $f(D)$ is considered to be defined on the \emph{whole set} of substochastic matrices, which contains both non-symmetric as well as symmetric substochastic matrices. We first find the maximum of $f(D)$ on this set, and then argue that there is a $D$ matrix that produces this maximum and satisfies the restriction expressed in Eq.~\eqref{eq:RDefined}.

As the coefficient of each $D_{ab}$ is non-negative, the maximum of  Eq.~\eqref{eq:expValueInType1} takes place on doubly-stochastic $D$ matrices, ($\sum_b D_{ab}=\sum_b D_{ba}=1$). The set of doubly-stochastic matrices is a convex set and the maximum of $f(D)$ is attained at its extreme points. By Birkhoff's theorem \cite{HornMatrixAnalysis2012a}, the extreme points of the set of doubly-stochastic matrices are permutation matrices, i.e., matrices where each row and each column has a single non-zero entry which is $1$. Consequently, $f(D)$ is maximized by a permutation matrix $D$. 

The cycle structure of the permutation represented by the maximizing $D$ matrix can be determined as follows. If the permutation has an $\ell$ cycle with $\ell\geq2$, $f(D)$ depends on the associated $m_a$ values by terms of the form
\begin{equation}
 f(D)=\cdots+\frac{1}{2}\left( m_1m_2+m_2m_3+\cdots+m_{\ell}m_1\right)+\cdots
\end{equation}
where indices are re-labeled for convenience. It will be shown that if $\ell\geq3$, then $f(D)$ can be made larger by reducing the cycle length of the permutation. For $\ell\geq5$, the $\ell$-cycle can be replaced by an $(\ell-2)$-cycle by conjoining two alternating label pairs, which makes $f(D)$ larger because
\begin{widetext}
\begin{equation}
 m_\ell m_1+m_1m_2+m_2m_3+m_3m_4+m_4m_5 < m_\ell(m_1+m_3)+(m_1+m_3)(m_2+m_4)+(m_2+m_4)m_5 ~.
\end{equation}
\end{widetext}
Similarly, a 4-cycle can be replaced by a 2-cycle because
\begin{equation}
\frac{1}{2}(m_1m_2+m_2m_3+m_3m_4+m_4m_1)  < (m_1+m_3)(m_2+m_4) ~.
\end{equation}
And a 3-cycle can be replaced by a 2-cycle, because, if $m_1\geq m_2,m_3$, then
\begin{equation}
 \frac{1}{2} \left( m_1m_2+m_2m_3+m_3m_1\right)< m_1(m_2+m_3)~.
\end{equation}
It is left to the reader to show that a 1-cycle can be replaced by a 2-cycle without decreasing $f(D)$. This establishes that the permutation associated with $D$ has only $2$-cycles. Finally, it is straightforward to see that converting several different 2-cycles into a single 2-cycle will make $f(D)$ larger. As a result, the maximum is obtained for $p=2$ single-particle orbitals and the corresponding $D$ (and consequently $R$) matrix is
\begin{equation}
 D=R=\left[\begin{array}{cc} 0 & 1 \\ 1& 0\end{array} \right]~~.
\end{equation}
It can be seen that these conditions are satisfied by a pair of mutually orthogonal single-particle states which are conjugates of each other by the anti-linear relation given above, i.e., $\alpha_2=\alpha_1^*$. Obviously, $\alpha_1$ should be orthogonal to $\alpha_1^*$ and should be a superposition of only the first $r$ standard basis states. A simple example that satisfies the conditions is given by
\begin{align}
\alpha_1 &= \left(\frac{1}{\sqrt2},\frac{i}{\sqrt2},0,0,\ldots\right)\,, \\
\alpha_2 &= \left(\frac{1}{\sqrt2},\frac{-i}{\sqrt2},0,0,\ldots\right)\,.
\end{align}
Because of this, it can be seen that
\begin{align}
\Lambda^\mathrm{sep (type 1)}_{r,N} &= \max_{m_1+m_2=N} m_1m_2\\
  &=\left\{\begin{array}{ll}  \frac{N^2}{4} & \textrm{if } N \textrm{ is even} \\
  \frac{N^2-1}{4} &  \textrm{if } N \textrm{ is odd} \end{array}
\right.
\end{align}
Note that, as long as $r\geq2$, the value of $r$ does not change anything in the derivation above, because there are only $p=2$ single-particle states.

\bibliographystyle{apsrev4-1}

\bibliography{pairing_witnessing_arxiv}

\end{document}